\documentclass{caosp306}
\usepackage{natbib}
\usepackage{graphicx}   
\usepackage{amsmath}    
\usepackage{amssymb}    
\usepackage{multicol}   
\usepackage{bm}         
\usepackage{pdflscape}  
\usepackage{float}      
\usepackage{tipa}       
\usepackage{amsmath}
\usepackage{breqn}
\usepackage{subfig}
\usepackage{natbib}

\def\BibTeX{{\rm B\kern-.05em{\sc i\kern-.025em b}\kern-.08em
             T\kern-.1667em\lower.7ex\hbox{E}\kern-.125emX}}

\begin{document}

\author{
        Daley-Yates S. \inst{1}
      \and 
        Stevens I. R. \inst{1}   
       }

\title{Winds of Massive Magnetic Stars: Interacting Fields and Flow}

\institute{
           School of Physics and Astronomy, University of Birmingham,\\
           Edgbaston, Birmingham,\\
           B15 2TT, U.K.\\ 
           \email{sdaley@star.sr.bham.ac.uk}
          }

\date{March 8, 2003}

\maketitle

\begin{abstract}
We present results of 3D numerical simulations of magnetically confined, radiatively driven stellar winds of massive stars, conducted using the astrophysical MHD code Pluto, with a focus on understanding the rotational variability of radio and sub-mm emission. Radiative driving is implemented according to the Castor, Abbott and Klein theory of radiatively driven winds. Many magnetic massive stars posses a magnetic axis which is inclined with respect to the rotational axis. This misalignment leads to a complex wind structure as magnetic confinement, centrifugal acceleration and radiative driving act to channel the circumstellar plasma into a warped disk whose observable properties should be apparent in multiple wavelengths.  This structure is analysed to calculate free-free thermal radio emission and determine the characteristic intensity maps and radio light curves.
%
\end{abstract}

\section{Introduction}
\label{sec:intro}
Radio emission from massive stars has historically been the subject of considerable interest \citep{Braes1972, Wright1974, Wright1975, Cohen1975}. This is, in part, due to the deviation from the expected results from Plank curve calculations of their millimetre and radio fluxes, which are substantially higher than expected. The additional emission is thought to be due to free-free interactions between charged species in the stellar wind. The spectral index of the continuum radio flux of a massive star is an indirect measure of its wind mass-loss \citep{Wright1975, Daley-Yates2016}. An accurate model for calculating stellar wind mass-loss needs to take account of any factor which can influence this spectral index. Perturbations that leads to non-spherical symmetry such as binary interactions \citep{Pittard2010b} or stellar magnetic fields, the subject of this proceeding, can affect this. The magnetic confinement of massive star winds has also been proposed as an explanation for time dependant H$\alpha$ emission \citep{Townsend2005b} as well as confined wind shocks which result in X-ray emission \citep{Ud-Doula2016}.

 \cite{ud-Doula2013c} presented the first 3D MHD simulations of a massive star wind, studying a highly symmetric configuration, with stellar rotational and magnetic field axes aligned with the z-axis. The magnetic field results in a disk structure in the equatorial region of the star. They also concluded that 2D time averaged simulation agree well with 3D spacially averaged results. Papers focusing on inclined magnetic fields of massive stars are absent from the literature. Inclined fields lead to the breaking of both symmetry and the disk like structure which develops in the magnetosphere of the star. This symmetry breaking is investigated here. The following section describes the parameters of the simulated star.

\section{Simulations}

The simulated star is based on model \textit{S3} from \cite{Daley-Yates2016}, which is summarised in Table \ref{tab:params}. Parameters were chosen to provide a typical massive star wind, with a dipolar magnetic field centred at the origin. 

\begin{table}
\caption{Parameters of the simulated star. \label{tab:params}}
\centering
\begin{tabular}{ccc}
\hline
Parameter & Symbol & Value \\
\hline
Stellar radius &$R_{*}$ & 9 $R_{\sun}$ \\
Stellar mass &$M_{*}$ & 27 $M_{\sun}$  \\ 
Equatorial magnetic field strength & $B_{eq}$ & 300 $G$ \\
Mass-loss rate &$\dot{M}_{B=0}$ & $3.2 \times 10^{-8}$ $M_{\sun} yr^{-1}$   \\
Wind terminal velocity & $V_{\infty}$ & 3000 $km s^{-1}$ \\
Escape velocity &$V_{\mathrm{esc}}$ & 1000 $km s^{-1}$ \\
Rotational rate as fraction of critical &$\omega$ & 0.2 $\omega_{\mathrm{crit}}$ \\ 
CAK force multiplier &$\alpha$ & 0.6 \\
Velocity law exponent &$\beta$ & 0.8 \\
\hline
\end{tabular}
\end{table}

The public code PLUTO (version 4.2) was used to solve the MHD equations \citep{Mignone2007}. Our simulations were conducted in spherical polar coordinates and covered an extent of $r~\in~[1 R_{*}, 40 R_{*}]$ in 300 cells, $\theta~\in~[0, \pi]$ in 120 cells and $\phi~\in~ [0, 2 \pi]$ in 120 cells. This region was divided into a mesh with resolution in the $r$ direction increasing from $\Delta r_{1}~\approx~0.00024 \ \mathrm{R_{\ast}}$ to $\Delta r_{300}~\approx~1.0 \ \mathrm{R_{\ast}}$. Both the $\theta$ and $\phi$ directions had the same resolution of $\Delta \theta_{j}~\approx~0.026 \ \mathrm{radians}$ and $\Delta \phi_{k}~\approx~0.026 \ \mathrm{radians}$ respectively.

Winds of massive stars are driven by absorption in emission lines, as described by \citet*{Castor1975} (CAK theory hereafter). According to CAK theory, wind material accelerates as
\begin{equation}
  g_{\mathrm{L}} \propto \left( \frac{dr/dv}{\rho} \right)^{\alpha} ,
\end{equation}
where $dr/dv$ is the radial velocity gradient. $g_{\mathrm{L}}$ is incorporated into the simulations as a source term in the momentum component of the MHD equations.

Thermal radio emission was calculated accurding to \cite{Daley-Yates2016} with the spectral flux density given by
\begin{equation}
  S_{\nu} = \int_{0}^{\infty} \frac{I(\nu, T)}{D^{2}} \mathrm{d}V.
\label{eq:flux_full}
\end{equation}
$I(\nu, T)$ is the intensity of radio emission at frequency $\nu$, $T$ is the temperature, $D$ is the distance from the star to the observer and $V$ is the computational volume.

\section{Results and Conclusions}

\begin{figure}
\centerline{\includegraphics[width=0.8\textwidth,clip=]{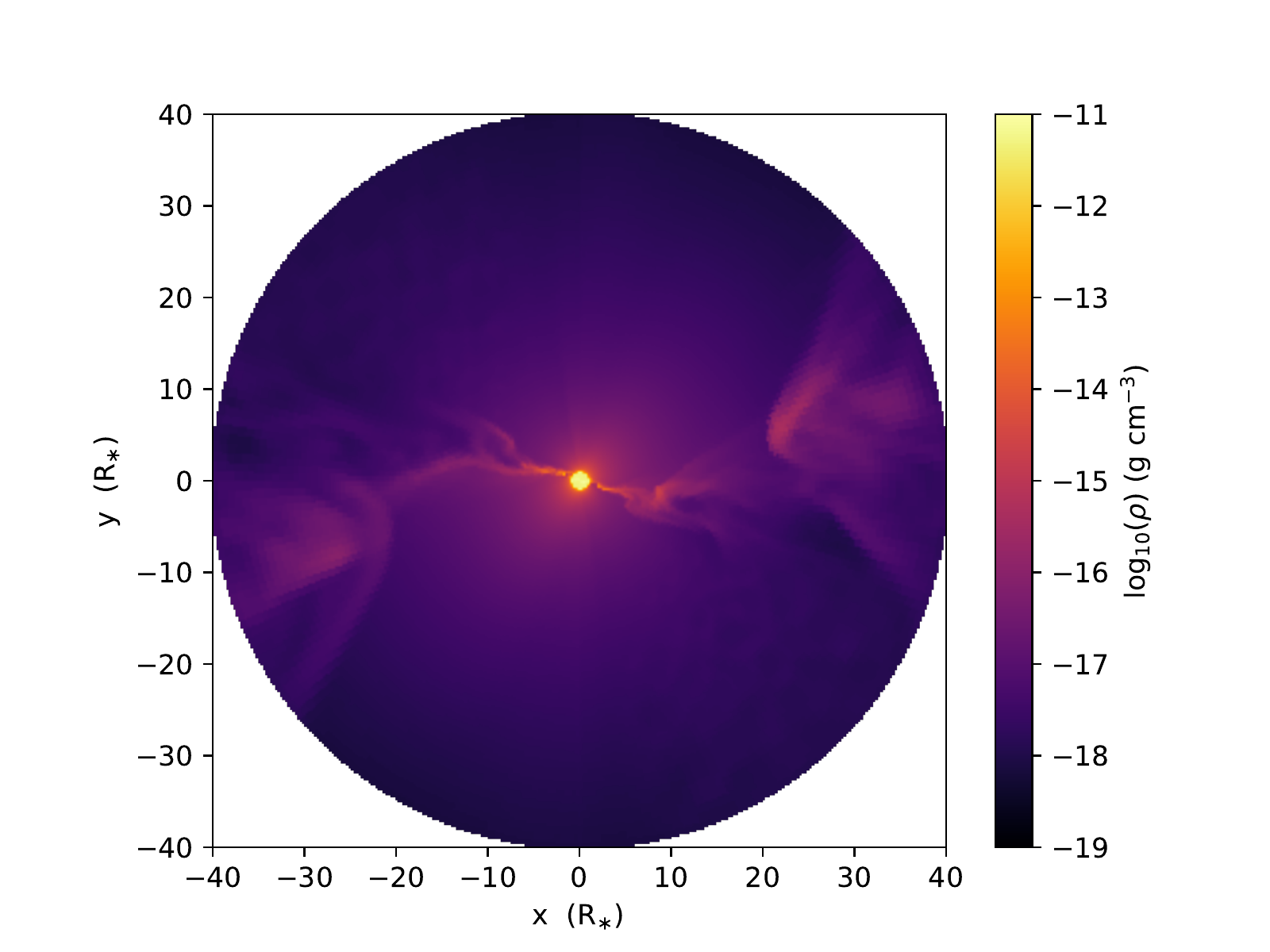}}
\caption{Density structure with the simulated star is at the centre and disk structure extending to roughly 10 $R_{*}$ before breaking up into diffuse clumps which expand as they move radially outwards.}
\label{fig:dens}
\end{figure}

The simulations were evolved for a total of 1 Ms in order to reach quasi-steady state. Fig. \ref{fig:dens} shows the resulting density structure. The star is at the centre with a disk structure extending into the equatorial region. The inclination of the dipole magnetic field, clockwise by 30$^{\circ}$, warps the disk until roughly 10 $R_{*}$ where it breaks up into diffuse clumps, which expand as they move radially outwards. Clumping behaviour is common in massive star winds, however the mechanism responsible is due to the intricacies of radiative line driving, which is not captured in these simulations. The interested reader is directed towards \cite{Owocki1999} for an in-depth analysis.

\begin{figure}
\begin{tabular}{cc}
\includegraphics[width=0.49\textwidth,clip=]{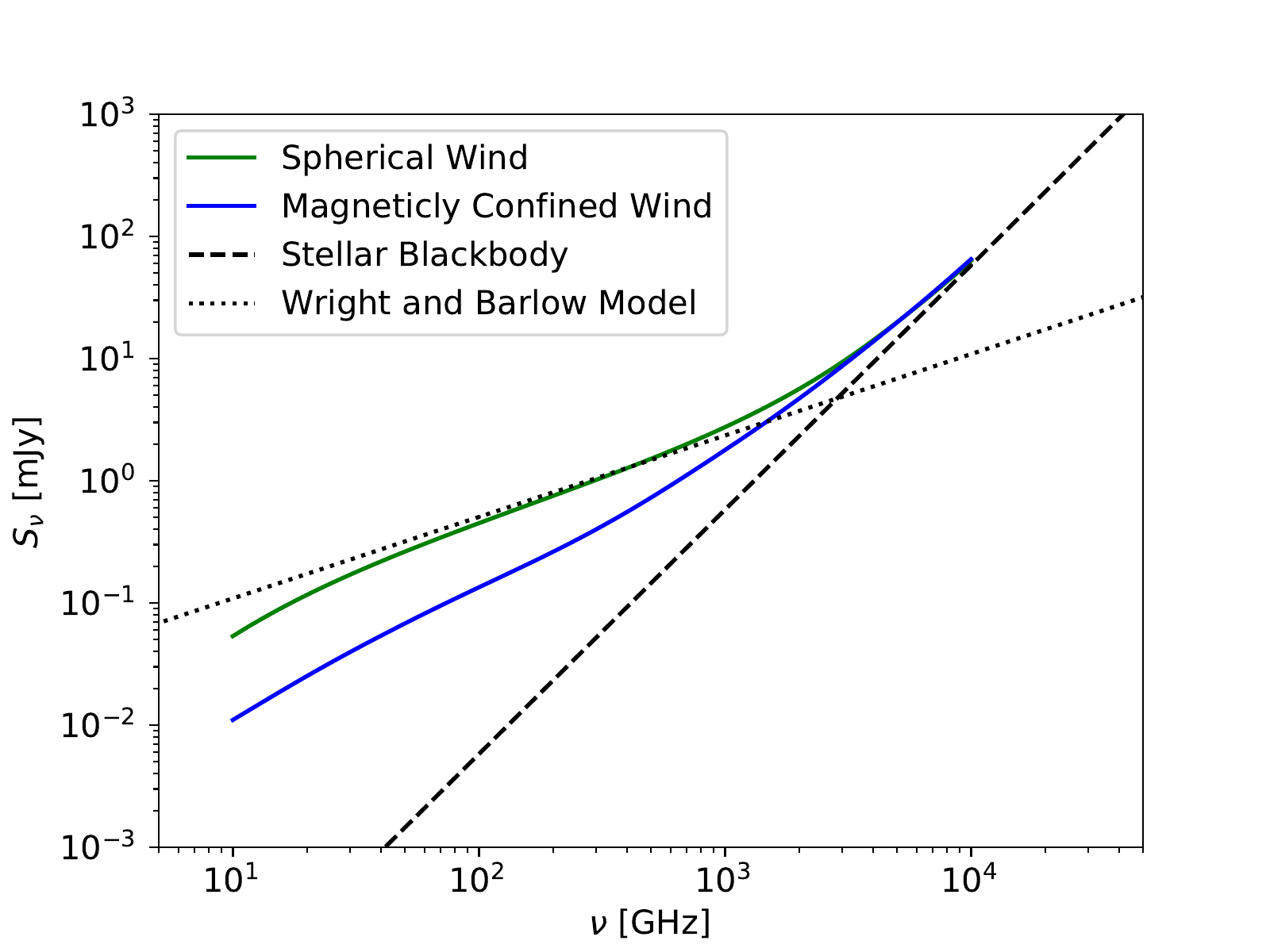} & 
\includegraphics[width=0.49\textwidth,clip=]{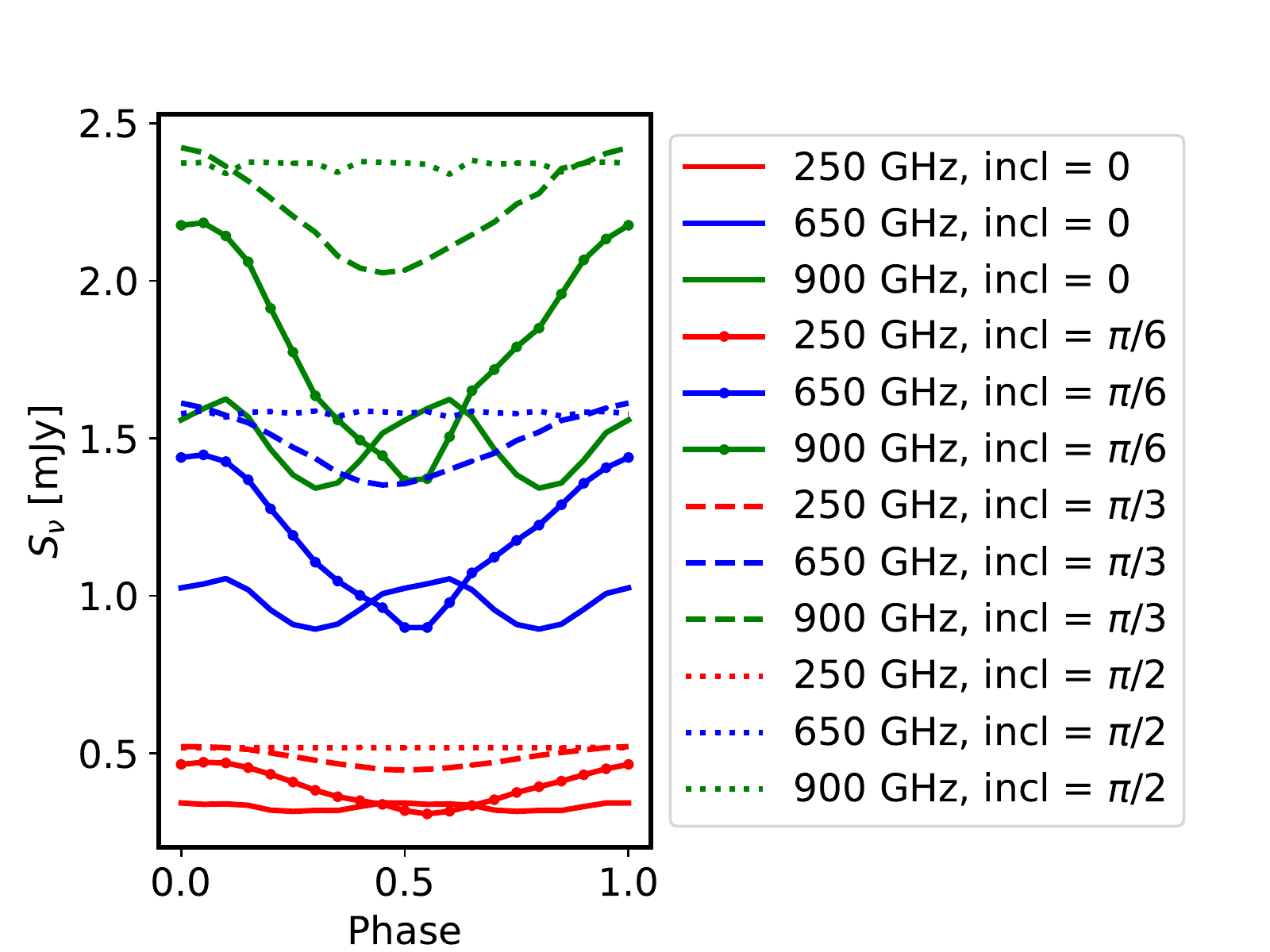} \\
\end{tabular}
\caption{Left: rotational modulation of the radio emission at discrete frequencies and observer inclination angles. Right: radio continuum spectrum for the three models described in the text. The black dashed line is the stellar Planck curve the other lines are described in the text.}
\label{fig:cont}
\end{figure}

Synthetic radio spectra for three models are presented on the left hand image of Fig. \ref{fig:cont}. These models are used to compare spherically symmetric (both analytic \citep{Wright1975} and numerical) and the magnetically confined wind results from Fig. \ref{fig:dens}. Both spherically symmetric calculations result in good agreement, only deviating at low frequencies, due to the finite size of the computational box, as low frequency emission originates at larger radii. The spectrum of the magnetically confined wind exhibits strong deviation at all radii except at frequencies close to the Planck curve, where emission is dominated by the stellar surface. This reduction in emission is due to the confinement of the wind material into a dens structured disk in a significant departure from spherically symmetry. The free-free optical depth is proportional to the density squared, $\tau_{\mathrm{ff}}~\propto~\rho^{2}$. Material otherwise observable is obscured and increases the spectral index of the emission, leading to a decrease in emission at frequencies below the Planck curve. 

Time dependent emission is illustrated in the right hand side of Fig. \ref{eq:flux_full}. Four angles of inclination between $0$ and $\pi/2$ are explored for 250 GHz, 650 GHz and 900 GHz. Both higher frequencies and observing inclination angles lead to greater emission. This is explained by considering that emission from the far side of the disk is obstructed when viewed side on. When viewed top down, much less obstruction takes place and a greater flux reaches the observer. 

Both the radio spectrum and light curves show deviations from the behaviour of spherical emission. This demonstrates that it is insufficient to simply know the emission at a given time. A full light curve, at numerous frequencies, is needed to fully understand the thermal emission from magnetically confined massive star winds.  

\bibliography{/Users/simon/Work/Reading/Reference_papers/Bibtex/library}

\end{document}